\documentclass[twocolumn,showpacs,preprintnumbers,amsmath,amssymb]{revtex4}
\usepackage{amssymb}
\usepackage[dvips]{graphicx}
\begin{document}
\textbf{Intrinsic tunneling or Joule heating?}

In a recent letter [1] Yurgens {\it et al.} reported superconducting $\Delta_s$ and normal  $\Delta_p$ (pseudo)gaps in the c-axis differential conductance of Bi2201-La mesas and ruled out a precursor Cooper pair scenario as an explanation of $\Delta_p$, while mechanisms based on the Van Hove singularity in the density of states, or on the
resonant tunnelling between $CuO_2$ planes, were outlined as possible explanations of the normal state pseudo-gap. I will show that a Joule heating of mesas may cause $I(V)$ nonlinearities similar to those observed in Ref.[1].

This approach provides a natural explanation of another finding of Yurgens {\it et al.} [1], the surprising similarity between the `intrinsic Josephson effect' (IJE) characteristics of Bi2201 and Bi2212 despite of a threefold difference in their critical temperatures. Moreover, similar (to those applied to Bi2212) levels of dissipation (estimated as $VI/A$) are required to achieve the characteristic features of IJE-spectra of Bi2201 at the same temperature, 4.2-4.5K (Fig.1,2a [1])  namely, 0.03-0.1, 0.6-3.9, and 2.5-15.6$kW/cm^2$ for the end of `brush', $\Delta_s$ and $\Delta_p$ respectively (the scatter reflects an uncertainty in the mesa area, $A$, in [1]). The relevance of heating in Bi2212 is experimentally confirmed by different methods; in particular, significant overheating of the mesa's surface was reported by [2,3] even in the `brush'-like part of I(V). I believe that the character of Joule heating of a sample with typical c-axis R(T) dependence, may be responsible for the similarities reported in [1].

 Here I will show that the `tunnelling characteristics' from Fig.4 [1] could be reproduced qualitatively {\it and} quantitatively using the experimental out-of-plane normal state resistance $R_c(T)$ (inset in Fig.4 of Ref.[1]) and assuming that the heating of the mesa caused by the Joule dissipation is the {\it only} reason for effects observed at high bias. Using  {\it Newton's Law of Cooling} (1701), the temperature of a thin mesa is given by
\begin{equation}
T=T_{0}+ IV/(Ah),
\end{equation}
where $T_0$ is the temperature of a coolant medium (liquid or gas) and $h$ is the heat transfer coefficient. According to Eq.(1), monitoring I(V) at certain base temperature $T_0$ results in a sample temperature rise that entails non-linearity in $I=V/R_c(T)$. Thus constructed $dI/dV$ curves resemble those presented in fig.4 [1]. This similarity allows for an estimate of the inverse heat transfer coefficient $(hA)^{-1}\simeq62.5K/mW$ and overheating at high bias, $\sim80K$ for $T_0=200K$; the set of curves accounting for this coefficient is shown in Fig.1. As it is clearly seen from Fig.1, the `heating' spectra taken  at $T_0<T^*$ reveal a `pseudo-gap' which disappears entirely as soon as $T_0>T^*$. This result shows that Eq.(1) provides a natural explanation of some of the puzzling findings of [1].
\begin{figure}
\begin{center}
\includegraphics[angle=-0,width=0.47\textwidth]{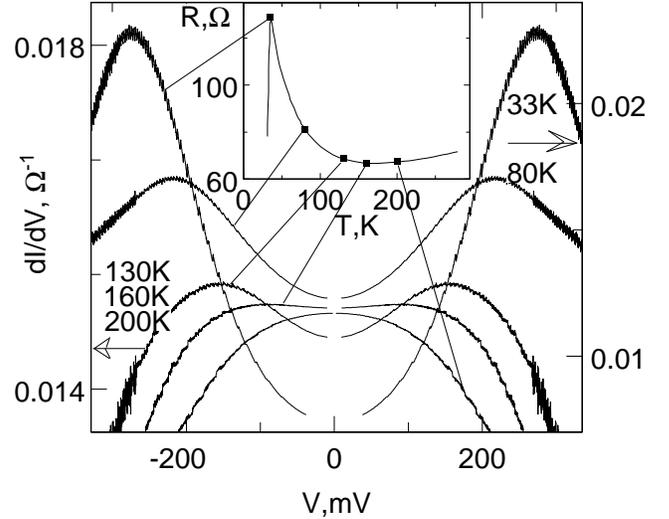}
\vskip -1mm
\caption{$dI/dV$ obtained from $R_c(T)$ of [1] (shown in the insert) assuming Joule heating origin of I(V) nonlinearities.}  \end{center}
\end{figure} 

To conclude, the nonlinear I(V) characteristics observed by [1] in Bi2201 mesas are related to the temperature dependence of the {\it normal} state c-axis resistance, rather than to the `Josephson tunnelling' between the
planes. As far as the unusual $R_c(T)$, itself  is concerned, a few models were proposed. In particular, the $R_c(T)$ explanation in the framework of the bipolaron model of cuprates [4] was supported experimentally [5]. In this model the temperature-independent-pseudogap is half of the binding energy of bipolarons.  Although only the normal state data were considered here, there is no doubt that heating plays an even more important role at low temperatures, so that heating issues have to be accounted for in the analysis of the low temperature data also.

The financial support of the Leverhulme Trust (F/00261/H) is gratefully acknowledged.

V.N.Zavaritsky 

Department of Physics, Loughborough University, Loughborough LE11 3TU, United Kingdom.

\medskip
\small
\noindent
[1]  A.Yurgens et al., Phys. Rev. Lett. {\bf 90}, 147005,(2003).

\noindent[2] C.E.Gough et al., cond-mat/0001365(2000).

\noindent[3]  V.N.Zavaritsky, J Superconductivity (US) {\bf15}, 567 (2002).

\noindent[4] A.S. Alexandrov et al.,  Phys. Rev. Lett. {\bf 77}, 4796 (1996).
  
\noindent[5]   V.N.Zavaritsky {\it et al.}, Europhys. Lett, {\bf 51}, 334, (2000);

 V.N. Zverev and D.V. Shovkun, JETP Lett. {\bf 72}, 73 (2000).
\end{document}